\documentclass[american,a4paper,runningheads,envcountsect]{llncs}

\usepackage{babel}
\usepackage[utf8]{inputenc}
\usepackage[T1]{fontenc}

\usepackage{enumitem}
\usepackage{amsmath,amsthm,amssymb}
\usepackage{mathtools}
\usepackage{mathrsfs}
\usepackage{fca}
\usepackage{dsfont} 

\usepackage[colorlinks,citecolor=blue,urlcolor=black,hidelinks,linktocpage]{hyperref}
\usepackage[all]{hypcap}
\usepackage{cleveref}
\let\cref\Cref

\theoremstyle{definition}

\theoremstyle{remark}

\renewcommand{\epsilon}{\varepsilon}
\renewcommand{\phi}{\varphi}

\usepackage[shrink=25,babel,kerning=true]{microtype}
\usepackage{csquotes}
\usepackage{booktabs}
\usepackage[subtle]{savetrees}
\usepackage{todonotes}
\presetkeys{todonotes}{color=blue!5}{}

\usepackage[backend=bibtex,style=numeric-comp,doi=false,isbn=false,%
url=false]{biblatex}
\newcommand\blfootnote[1]{%
  \begingroup
  \renewcommand\thefootnote{}\footnote{#1}%
  \addtocounter{footnote}{-1}%
  \endgroup
}

\begin{document}

\title{\texttt{DimDraw} -- A novel tool for drawing concept lattices}

\author{Dominik Dürrschnabel \and Tom Hanika \and Gerd Stumme}

\date{\today}

\institute{%
  Knowledge \& Data Engineering Group,
  University of Kassel, Germany\\[1ex]
  \email{$\{$duerrschnabel,hanika,stumme$\}$@cs.uni-kassel.de}
}
\maketitle

\blfootnote{Authors are given in alphabetical order.
  No priority in authorship is implied.}

\begin{abstract}
  Concept lattice drawings are an important tool to visualize complex relations
  in data in a simple manner to human readers.  Many attempts were made to
  transfer classical graph drawing approaches to order diagrams. Although those
  methods are satisfying for some lattices they unfortunately perform poorly in
  general. In this work we present a novel tool to draw concept
  lattices that is purely motivated by the order structure.
\end{abstract}


\section{Introduction} 
Line diagrams are a great tool for interpreting data ordered through formal
concept analysis. In such diagrams every concept is visualized as a dot on the
plane and the covering order relations are visualized as straight lines that are
not allowed to touch other concept dots. Additional to these strong
conditions, there are several soft conditions to improve the readability of
diagrams for a human reader. For example, to minimize the number of crossing lines
or to minimize the number of different slopes. Another desirable condition is to
draw as many chains as possible on straight lines. Lastly, the distance of dots
to (non-incident) lines should be maximized. Experience shows that in order to obtain a (human) readable drawing one has to balance
those criteria. Based on this idea, there are algorithms and tools that work well on order
diagrams. However, such automated drawings usually cannot compete to those
created manually by an experienced human. Since such an expert is often not
available, the task for creating a suitable (and easily readable) line diagram for a
given concept lattice is a challenging task.

Here we step in with our novel tool \texttt{DimDraw}. We claim that it produces
drawings that come close to the quality of hand-drawn line diagrams. In contrast
to prior approaches it tries not only to optimize on some set of criteria but also
employs the order structure of the concept lattice itself. 
More precisely, our tool computes a family of linear orders, called
\emph{realizer}, that entails the to-be-drawn lattice order as
its intersection. From this family we derive coordinates for our lattice drawing. In
fact, our tool
does not only work on concept lattices but also on arbitrary orders.

\subsubsection{Related Work and Basics from Formal Concept Analysis.}
We rely on notations from~\cite{fca-book}, i.e., $(G,M,I)$ is a \emph{formal
  context} with sets $G$ and $M$, and $I\subseteq G\times M$. Furthermore,
$\cdot’\colon \mathcal{P}(G)\to\mathcal{P}(M),A\mapsto A’\coloneqq\{m\in
M\mid\forall g\in A: (g,m)\in I\}$ and
$\cdot’\colon\mathcal{P}(M)\to\mathcal{P}(G),B\mapsto B’\coloneqq\{g\in G\mid
\forall m\in B:(g,m)\in I\}$ are \emph{derivation operators} enabling
$\mathfrak{B}(G,M,I)=\{(A,B)\subseteq\mathcal{P}(G)\times\mathcal{P}(M)\mid
A’=B\wedge B’=A\}$, the set of all \emph{formal concepts}. This set can be
ordered through the inclusion order on $G$.  Various approaches for drawing line
diagrams can be found: The author in~\cite{Wille1989} employs an order approach
which lays the foundation for our work. In~\cite{freese2004automated} a rank
function and a forced directed approach was used and the author
in~\cite{ganteradd} focuses on additive diagrams.

\section{Order Dimension Approach}
\label{sec:order-dimens-appr}
Before introducing the algorithm behind \texttt{DimDraw} we need to recollect
some basic notions on order dimension.  Let $(X,\leq_P)$ be an ordered set,
i.e., $\leq_{P}\subseteq X\times X$ is an order relation.  Two elements $x,y\in
X$ are called \emph{incomparable} in $P$, if neither $x\leq_P y$ nor $y\leq_Px$,
otherwise they are called \emph{comparable}. If $(X,\leq_L)$ is an ordered set
with $\leq_P\subseteq \leq_L$, such that no two elements are incomparable in
$\leq_L$, then $\leq_L$ is called a \emph{linear extension} of $\leq_P$. The
\emph{intersection} of two order relations $\leq_{L_1}$ and $\leq_{L_2}$ on $X$
is also an order relation on $X$.  Let $\mathscr{L}$ be a set of linear
extensions of $\leq_P$ with $\bigcap_{\leq_L \in
  \mathscr{L}}\leq_L=\leq_P$. Then $\mathscr{L}$ is called a \emph{realizer} of
$\leq_P$. The \emph{order dimension} of an order relation $\leq_{P}$ is the
cardinality of a minimal realizer of $\leq_{P}$.

A \emph{Ferrers relation}  is a relation
$F\subseteq G \times M$, such that for all $g,h\in G$ and $m,n \in M$ the
following condition holds: $(g,m)\in F$ and ${(h,n)\in F \Rightarrow (g,n)\in F}$
or $(h,m)\in F$. Following~\cite[Proposition 103]{fca-book} we know that
$F\subseteq G\times M$ is a Ferrers relation if and only if
$\mathfrak{B}(G,M,F)$ is a chain, i.e., all elements of $\mathfrak{B}(G,M,F)$
are pairwise comparable. Those chains are essential for our drawing
approach. The \emph{Ferrers dimension} of a formal context is the smallest
number $k$, such that there exists a set of $k$ Ferrers relations with their
intersection being $I$. The Ferrers dimension of a context is equal to the order
dimension of its concept lattice, see~\cite[Theorem 46]{fca-book}. We further
know that the order dimension of $\mathcal{B}(G,M,I)$ for some context $(G,M,I)$
is at most $d$, if there are Ferrers relations $F_1,\dotsc,F_d\subseteq G \times
M$ with $G \times M \backslash I = \bigcup\nolimits_{i=1}^d F_i$.  This fact
gives a handy way to compute the order dimension of a concept lattice. One has
to cover all empty cells of the cross table of a concept with Ferrers
relations. Note that those do not have to be disjoint. Unfortunately deciding
the order dimension and the Ferrers dimension is $\mathcal{NP}$-complete if the
dimensions is three or higher.

\subsubsection*{Computing Coordinates using a Realizer.}
\label{sec:comp-coord-using}
First we have to compute a minimal realizer of the order relation of the concept
lattice. This can be done with an algorithm described in
\cite{yanez1999poset}. Note that the algorithm described there is only exact for
certain ordered sets. Alternatively one can use the Ferrers dimension: Try to cover
all the empty spaces in a cross-table by as few Ferrers relations as
possible. Then take the inverse Ferrers relations, compute the chains given by
these relations and extend them to cover every element of $\mathfrak{B}(G,M,I)$.

Given a linear extension of a lattice, let the \emph{position} of each concept
in the linear extension be the cardinality of the set of sub-concepts. Embed the
concept lattice into $\mathbb{R}^d$ with $d$ being its order dimension. The
coordinates of concept $C$ are given as $(c_1,\ldots,c_d)$, such that $c_i$ is
the position of concept $C$ in the linear extension $L_{i}$. Note that for each
pair of concepts $X,Y$ it holds that $X$ is a sub-concept of $Y$, if and only
if, for the coordinates $X=(x_1,\ldots,x_d)$ and $Y=(y_1,\ldots,y_d)$, it holds that
$x_i\leq y_i$ for all $1\leq i\leq d$.  By rotating the resulting embedding we
obtain a valid embedding of the concept lattice in $\mathbb{R}^d$. We have to
carefully project this embedding to the plane to obtain our drawing. The theory
we developed for the projection is very extensive and will be presented in a
later work due to space constraints.

\subsubsection{Example.}

Take the ``Life in water'' context from \cite{fca-book}, and consider the three Ferrers-relations that cover all free spaces in the cross table:

\begin{center}
   \hspace*{-.7cm} {\footnotesize
    \scalebox{.7}{
    \begin{cxt}
      \cxtName{}
      \att{$a$}
      \att{$b$}
      \att{$c$}
      \att{$d$}
      \att{$e$}
      \att{$f$}
      \att{$g$}
      \att{$h$}
      \att{$i$}
      \obj{x x . . . . x . .}{$1$}
      \obj{x x . . . . x x .}{$2$}
      \obj{x x x . . . x x .}{$3$}
      \obj{x . x . . . x x x}{$4$}
      \obj{x x . x . x . . .}{$5$}
      \obj{x x x x . x . . .}{$6$}
      \obj{x . x x x . . . .}{$7$}
      \obj{x . x x . x . . .}{$8$}
    \end{cxt}
    \begin{cxt}
      \cxtName{}
      \att{$a$}
      \att{$b$}
      \att{$c$}
      \att{$d$}
      \att{$e$}
      \att{$f$}
      \att{$g$}
      \att{$h$}
      \att{$i$}
      \obj{x x b . b b x b b}{$1$}
      \obj{x x b . b . x x b}{$2$}
      \obj{x x x . b . x x b}{$3$}
      \obj{x . x . b . x x x}{$4$}
      \obj{x x b x b x . . b}{$5$}
      \obj{x x x x b x . . b}{$6$}
      \obj{x . x x x . . . .}{$7$}
      \obj{x . x x b x . . b}{$8$}
    \end{cxt}
    \begin{cxt}
      \cxtName{}
      \att{$a$}
      \att{$b$}
      \att{$c$}
      \att{$d$}
      \att{$e$}
      \att{$f$}
      \att{$g$}
      \att{$h$}
      \att{$i$}
      \obj{x x . . . . x . u}{$1$}
      \obj{x x . . . . x x u}{$2$}
      \obj{x x x . . . x x .}{$3$}
      \obj{x . x . . . x x x}{$4$}
      \obj{x x . x . x u u u}{$5$}
      \obj{x x x x . x u u u}{$6$}
      \obj{x u x x x u u u u}{$7$}
      \obj{x u x x . x u u u}{$8$}
    \end{cxt}
    \begin{cxt}
      \cxtName{}
      \att{$a$}
      \att{$b$}
      \att{$c$}
      \att{$d$}
      \att{$e$}
      \att{$f$}
      \att{$g$}
      \att{$h$}
      \att{$i$}
      \obj{x x . d d d x . .}{$1$}
      \obj{x x . d d d x x .}{$2$}
      \obj{x x x d d d x x .}{$3$}
      \obj{x d x d d d x x x}{$4$}
      \obj{x x . x d x . . .}{$5$}
      \obj{x x x x . x . . .}{$6$}
      \obj{x . x x x . . . .}{$7$}
      \obj{x . x x d x . . .}{$8$}
    \end{cxt}}
    }
  \end{center}
\noindent  
These Ferrers relations give rise to the following (not unique) realizer:

{\scriptsize
\noindent
$\bullet$\ $S\leq_1 Q \leq_1 O \leq_1 R \leq_1 P \leq_1 N \leq_1 H \leq_1 L \leq_1 K \leq_1
C \leq_1 M \leq_1 I \leq_1 G \leq_1 F \leq_1 J \leq_1 E \leq_1 B \leq_1 D \leq_1
A$\\
\noindent
$\bullet$\ $S\leq_2 P \leq_2 O \leq_2 M \leq_2 J \leq_2 H \leq_2 F \leq_2 B \leq_2 R \leq_2
K \leq_2 I \leq_2 D \leq_2 N \leq_2 G \leq_2 Q \leq_2 L \leq_2 E \leq_2 C \leq_2
A$\\
\noindent
$\bullet$\ $S\leq_3 R \leq_3 Q \leq_3 N \leq_3 I \leq_3 L \leq_3 G \leq_3 E \leq_3 P \leq_3 M \leq_3 K \leq_3 J \leq_3 D \leq_3 O \leq_3 H \leq_3 F \leq_3 C \leq_3 B \leq_3 A$}
\smallskip

\begin{figure}[t]
  \centering\hfill
  \includegraphics[height=0.15\textheight]{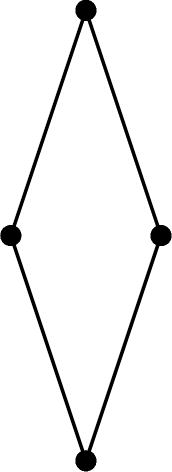}\hspace{1.5cm}
  \includegraphics[height=0.15\textheight]{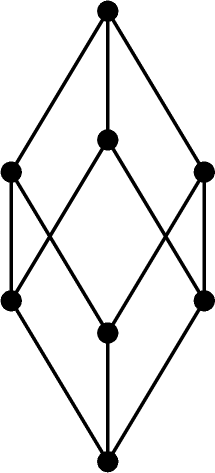}\hspace{1.5cm}
  \includegraphics[height=0.15\textheight]{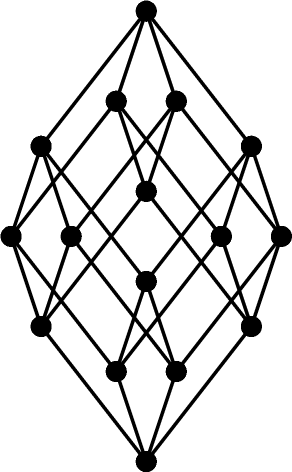}\hspace{1.5cm}\hspace{2.5em}
  \caption{Concept lattice line diagrams for the two, three and four dimensional scales.}
  \label{fig:dims}
\end{figure}

Note that this context is in fact 3-dimensional as the elements
$\{E,C,D,L,I,K\}$ form the ordered set $S_3$, which is well-known to be
3-dimensional.  Now we embed the concepts based on their positions in the linear
extensions as described above: {\footnotesize
  A(18,18,18), B(16,7,17), C(9,17,16), D(15,16,7), E(15,16,7), F(13,6,15),
  G(12,13,6), H(6,5,14), I(11,10,4), J(14,4,11), K(8,9,10), L(7,15,5),
  M(10,3,9), N(5,12,3), O(2,2,13), P(4,1,8), Q(1,14,2), R(3,8,1), and S(0,0,0).
} Now the remaining task is to project this into the plane, which is a research task
for itself. We obtain the following embedding (left) which is quite close  to the hand
drawn version (right):
\begin{center}\hfill
  \includegraphics[scale=1]{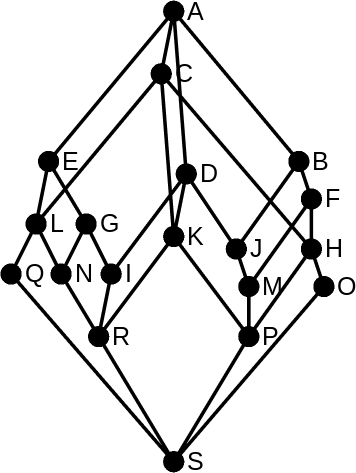}\hfill
  \includegraphics[scale=0.7]{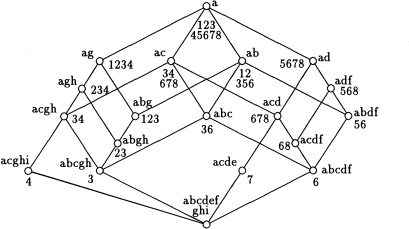}
\end{center}
As another example we present the \emph{contra nominal scales} on $n$ objects
and attributes, i.e., $([n],[n],\neq)$ where
$[n]\coloneqq\{1,2,\dotsc,n\}$. It is easy to see that these exhibit order dimension
$n$. We show line diagrams of the concept lattices computed by
\texttt{DimDraw} for the two, three, and four dimensional contra nominal scales
in~\cref{fig:dims}.

We conclude this section by comparing our tool to previous results
in~\cite{ganteradd}. There the author employed an approach to generate additive
line diagrams (left). We may note that our drawing (middle) is not additive,
however, it resembles the hand drawn additive diagram (right) more accurately.
\begin{center}\hfill
  \includegraphics[height=0.19\textheight]{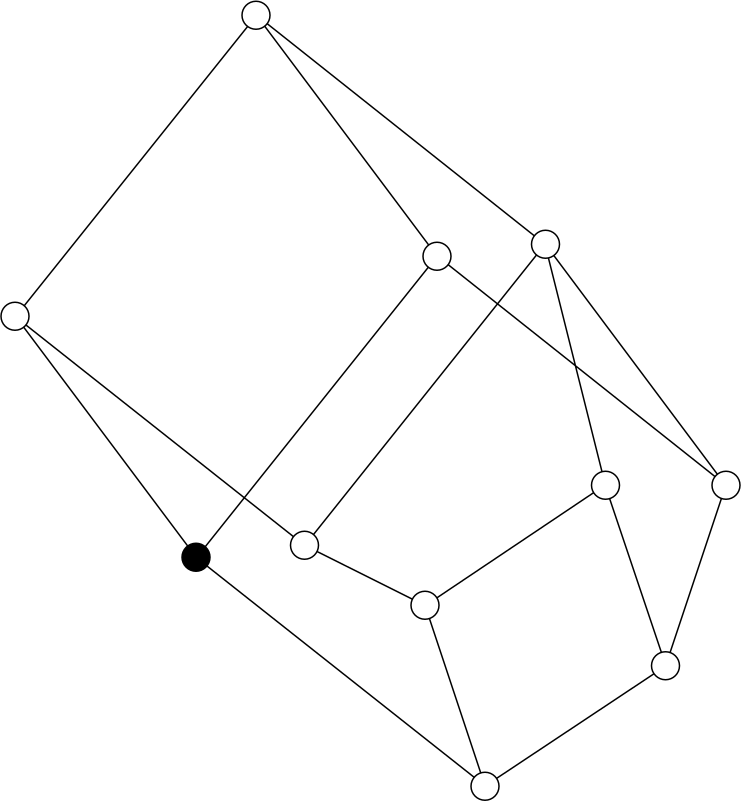}\hfill
  \includegraphics[height=0.19\textheight]{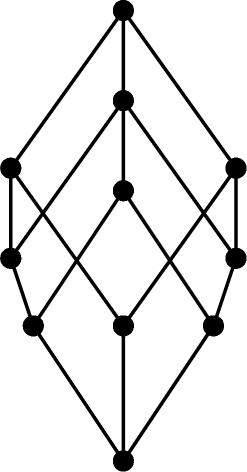}\hfill
  \includegraphics[height=0.19\textheight]{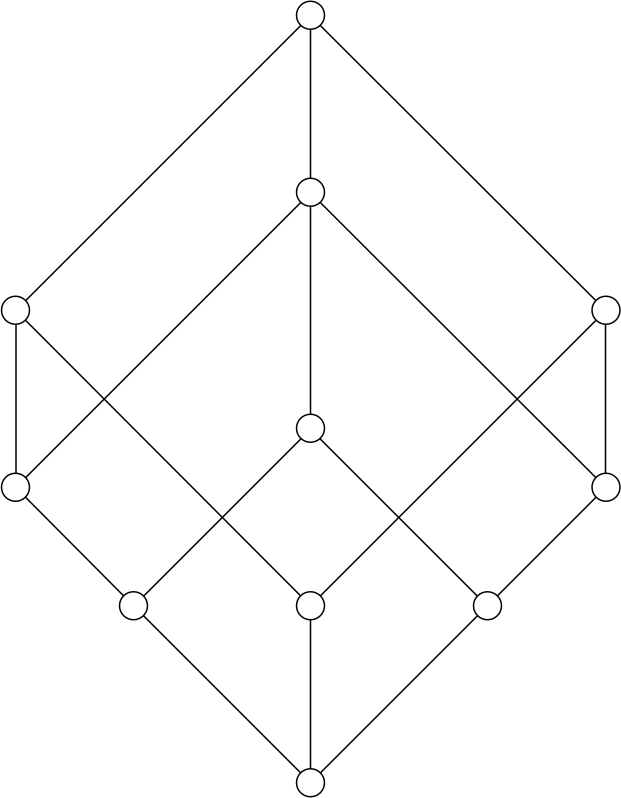}
\end{center}
\subsubsection{Outlook}
\label{sec:outlook}
In this work we demonstrated a novel approach for drawing lattice diagrams
automatically, which resulted in the work-in-progress software
\texttt{DimDraw}. This tool will be part of the upcoming version of
\texttt{conexp-clj} (\url{https://github.com/exot/conexp-clj}).  The ongoing
research for more sophisticated projection theories, as used by this tool, will
be extended in future publications.

\printbibliography

\end{document}